\documentclass[aps,prd,floatfix,groupedaddress,nofootinbib,twocolumn,superscriptaddress]{revtex4-1}

\usepackage{amsmath,amssymb,bm,color,dcolumn,mathrsfs,tabularx,graphicx}
\usepackage[hidelinks=true]{hyperref}
\hypersetup{colorlinks=true,linkcolor=blue,citecolor=blue,urlcolor=blue}
\usepackage{Macro} %to avoid def conflicts

\def\T{\Theta}

\def\cov{{\rm Cov}}
\newcommand{\tred}[1]{{#1}}
\def\mf#1{#1}

\begin{document}

% Title %
\title{Impact of nonlinear growth of the large-scale structure on CMB B-mode delensing}

% Authors %
\author{Toshiya Namikawa}
\affiliation{Leung Center for Cosmology and Particle Astrophysics, National Taiwan University, Taipei, 10617, Taiwan}
\author{Ryuichi Takahashi}
\affiliation{Faculty of Science and Technology, Hirosaki University, 3 Bunkyo-cho, Hirosaki, Aomori 036-8561, Japan}

% Date %
\date{\today}

% Abstract %
\begin{abstract}
We study the impact of the nonlinear growth of the large-scale structure (LSS) on the removal of the gravitational lensing effect (delensing) in cosmic microwave background (CMB) $B$ modes. The importance of the nonlinear growth of the LSS in the gravitational lensing analysis of CMB has been recently recognized by several works, while its impact on delensing is not yet explored. The delensing using mass-tracers such as galaxies and cosmic infrared background (CIB) could be also affected by the nonlinear growth. We find that the nonlinear growth of the LSS leads to $\sim 0.3\%$ corrections to $B$-mode spectrum after delensing with a high-$z$ mass tracer ($z_m\sim 2$) at $\l=1000$-$2000$. The off-diagonal correlation coefficients of the lensing $B$-mode template spectrum become significant for delensing with low-$z$ tracers ($z_m\alt 0.5$), but are negligible with high-$z$ tracers (such as CIB). On the other hand, the power spectrum covariance of the delensed $B$ mode is not significantly affected by the nonlinear growth of the LSS, and the delensing efficiency is not significantly changed even if we use low-$z$ tracers. The CMB $B$-mode internal delensing is also not significantly affected by the nonlinear growth. 
\end{abstract}

% Output Title/Abstract etc %
\maketitle

% Contents %

%////////////////////////////////////////////////////////////////////////////////////////////////////%
\section{Introduction} \label{intro}
%////////////////////////////////////////////////////////////////////////////////////////////////////%

% B-mode
The gravitational waves generated at the inflationary era induce the curl pattern in the cosmic microwave background (CMB) polarization map, the so-called $B$ mode (see e.g. \cite{Kamionkowski:2015yta} and references therein). The detection of the large-scale $B$ mode by the inflationary gravitational waves (IGWs) will open a new window to the early universe. The CMB experiments in the near future such as BICEP Array \cite{BICEPArray} and Simons Observatory \cite{SimonsObservatory} will be able to precisely measure the $B$ mode, and significantly improve the sensitivity to the IGWs. 

% delensing
The current upper bound on the amplitude of the IGWs (the so-called tensor-to-scalar ratio) is $r<0.07$ ($95\%$ C.L.) \cite{BKIV}. Since the lensing effect converts part of $E$ mode to $B$ mode, the removal of the lensing contamination (delensing) increases the signal-to-noise of the tiny IGW signals (e.g., \cite{Knox:2002pe,Kesden:2002ku,Seljak:2003pn,Hirata:2003ka,Pan:2017,Millea:2017}). The measured CMB lensing potential and/or tracers of the underlying matter distribution such as galaxies and cosmic infrared background (CIB) can be used to estimate the $E$-to-$B$ leakage by lensing, or to perform the inverse remapping of the lensed CMB map to reduce the lensing effect in the observed CMB map. 

% nonlinear growth
So far multiple works have discussed the feasibility of the delensing of CMB anisotropies in future experiments by the reconstruction of lensing potential internally (e.g., \cite{Smith:2010gu,Teng:2011xc,Sehgal:2016,Green:2016,Challinor:2018,Carron:2018}) and/or by employing external data such as CIB and radio galaxies (e.g., \cite{Simard:2015,Sherwin:2015,Namikawa:2015c}). Several works have already demonstrated the delensing using existing data \cite{Larsen:2016,SPTpol:delens,Carron:2017,P18:phi}. The impact of the nonlinear growth of the large-scale structure (LSS) is, however, not yet explored. Compared to the low-$z$ lensing, the lensing effect on CMB mostly comes from the gravitational potential at high redshifts, and is not significantly affected by the nonlinear growth of the underlying density fluctuations. However, some recent studies show that the nonlinear growth significantly affects the measurement of the lensing power spectrum \cite{Boehm:2016,Pratten:2016,Boehm:2018,Beck:2018}. Also, the lensing potential bi-spectrum and other higher-order statistics from the nonlinear growth of the LSS are detectable in the near future CMB experiments \cite{Namikawa:2016b,Pratten:2016,Liu:2016nfs,Namikawa:2018b}. 

% mass tracer delensing and nonlinear growth effect
The delensing using mass-tracers (such as galaxies and CIB) or reconstructed CMB lensing map could be also affected by the nonlinear growth of the LSS. The matter density fluctuations at late time of the universe are highly non-Gaussian. If mass tracers are used, the delensed $B$ mode by such non-Gaussian fields can have additional contributions from the higher-order correlations between the density fluctuations and lensing potential. As a result, the power spectrum of the delensed $B$ mode is modified. In addition, the covariance of the delensed $B$-mode spectrum is also affected by the non-Gaussianity of the density fluctuations at low redshifts. If the reconstructed lensing potential is used for delensing, the statistics of the delensed $B$ mode becomes more complicated. For example, the power spectrum of the delensed $B$-mode contains, at least, four- and six-point correlations of CMB anisotropies and has higher-order correlations of the lensing potential. 

% motivation
This paper addresses the impact of the nonlinear growth of the LSS on $B$-mode delensing by simulations. We compute full-sky maps of galaxy distributions and lensed CMB maps based on N-body simulations. The CMB deflection angle published by Ref.~\cite{Takahashi:2017} is also used for making the lensed CMB maps. The delensing is performed by computing the template of the lensing $B$ mode and subtracting it from the ``observed'' $B$ mode. We then compute the statistics of the delensed $B$ mode such as the angular power spectrum and its correlation coefficients. 

This paper is organized as follows. 
Section~\ref{delensing} recalls the basic methodology of the delensing. 
Section~\ref{sim} briefly describes our simulations used in our analysis. 
Section~\ref{result} presents our main results. 
Section~\ref{summary} summarizes our work. 
In addition to the nonlinear growth effect on delensing, appendix~\ref{appA} shows the survey boundary and sky projection effects in delensing. 

Throughout this paper, we adopt the flat $\Lambda$CDM (lambda cold dark matter) model consistent with the WMAP 9-yr result \cite{Hinshaw:2013}. The cosmological parameters are the matter density $\Omega_{\rm m} = 0.279$, the baryon density $\Omega_{\rm b} = 0.046$, the cosmological constant $\Omega_\Lambda = 0.721$, the Hubble parameter $h = 0.7$, the amplitude of density fluctuations $\sigma_8 = 0.82$, and the spectral index $n_{\rm s} = 0.97$.

%////////////////////////////////////////////////////////////////////////////////////////////////////%
\section{Delensing} \label{delensing}
%////////////////////////////////////////////////////////////////////////////////////////////////////%

Here we briefly summarize notations in this paper and methods of the delensing. 

%++++++++++++++++++++++++++++++++++++++++++++++++++++++++++++++++++++++++++++++++++++++++++++++++++++%
\subsection{Lensed CMB anisotropies}
%++++++++++++++++++++++++++++++++++++++++++++++++++++++++++++++++++++++++++++++++++++++++++++++++++++%

The lensing effect on the CMB anisotropies is well described by a remapping of the primary CMB anisotropies (e.g. \cite{Lewis:2006fu}). Denoting the primary CMB temperature and polarization anisotropies as $\T(\hatn)$ and $Q(\hatn)\pm\iu U(\hatn)$, the lensed CMB anisotropies in a direction $\hatn$ are obtained by 
%----------------------------------------------------------------------------------------------------%
\al{
	\tT(\hatn) &= \T(\hatn+\bm{d}(\hatn)) \,, \label{Eq:T:remap} \\
	[\tQ\pm\iu \tU](\hatn) &= [Q\pm\iu U](\hatn+\bm{d}(\hatn)) \,. \label{Eq:QU:remap}
}
%----------------------------------------------------------------------------------------------------%
Here, $\bm{d}(\hatn)$ is the deflection angle, and is usually decomposed into the gradient and curl modes, $\bn\grad$ and $(\star\bn)\curl$ \cite{Hirata:2003ka}, but the curl mode is negligible in the standard $\Lambda$CDM simulation \cite{Saga:2015,Takahashi:2017}. 

The harmonic coefficients of the scalar quantities, $\T$ and $\grad$, are obtained by the (spin-$0$) spherical harmonics transform, while the polarization parameters, $Q$ and $U$, are transformed to the following $E$ and $B$ modes by the spin-$2$ spherical harmonics, ${}_{\pm 2}Y_{\l m}$: 
%----------------------------------------------------------------------------------------------------%
\al{
	[E \pm \iu B ]_{\l m} = -\Int{}{\hatn}{} {}_{\pm 2} Y_{\l m}^*(\hatn) [Q\pm \iu U](\hatn)  \,. 
}
%----------------------------------------------------------------------------------------------------% 
Expanding \eq{Eq:QU:remap} up to the first order of the lensing potential, the lensing-induced B-mode is given by (e.g.~\cite{Hu:2000ee}) 
%----------------------------------------------------------------------------------------------------%
\al{
	B^{\rm lens}_{\l m} = -\iu\sum_{\l'm'}\sum_{LM} \Wjm{\l}{\l'}{L}{m}{m'}{M}\mC{S}_{\l\l'L} 
	    E_{\l'm'}\grad_{LM}
	\,, \label{Eq:Lensing-E-to-B}
}
%----------------------------------------------------------------------------------------------------%
where $\grad_{LM}$ is the harmonic coefficients of the lensing potential, and we ignore the primary $B$ mode and lensing curl mode. The quantity $\mC{S}_{\l\l'L}$ represents the mode coupling induced by lensing: 
%----------------------------------------------------------------------------------------------------%
\al{ 
	\mC{S}_{\l\l'L} 
		&= \sqrt{\frac{(2\l+1)(2\l'+1)(2L+1)}{16\pi}}\Wjm{\l}{\l'}{L}{2}{-2}{0}
	\notag \\
		&\qquad\times [-\l(\l+1)+\l'(\l'+1)+L(L+1)] 
	\,. \label{Eq:Spm}
} 
%----------------------------------------------------------------------------------------------------%
Here, the above quantity is unity if $\l+\l'+L$ is an odd integer and zero otherwise.

%++++++++++++++++++++++++++++++++++++++++++++++++++++++++++++++++++++++++++++++++++++++++++++++++++++%
\subsection{Delensing}
%++++++++++++++++++++++++++++++++++++++++++++++++++++++++++++++++++++++++++++++++++++++++++++++++++++%

A method to remove lensing contributions in observed $B$ modes is to make a template of the lensing $B$ modes (hereafter, lensing template) with a measured CMB lensing map and $E$ mode \cite{Seljak:2003pn,Smith:2010gu}. The lensing template up to first order of the CMB lensing potential is defined as \cite{Simard:2015,Sherwin:2015,Namikawa:2015c,Manzotti:2018}
%----------------------------------------------------------------------------------------------------%
\al{
	B^{\rm res}_{\l m} = -\iu\sum_{\l'm'}\sum_{LM} \Wjm{\l}{\l'}{L}{m}{m'}{M}\mC{S}_{\l\l'L} 
	    W^E_{\l'}\hE_{\l'm'} W^x_L\hx_{LM}
	\,, \label{Eq:delens-B}
}
%----------------------------------------------------------------------------------------------------%
where $\hE_{\l'm'}$ is the observed $E$ mode, $\hx_{LM}=x_{LM}+n_{LM}$ is an observed mass tracer which correlates with the CMB lensing potential. In this paper, $\hx$ is either the galaxy density fields or reconstructed CMB lensing potential. The quantity $W^E_\l$ is the $E$-mode Wiener filter defined as $W^E_\l=C_\l^{\rm EE}/\hCEE_\l$, where $C_\l^{\rm EE}$ and $\hCEE_\l$ denote the angular power spectra of $E$ and $\hE$, respectively. Similarly, the filter for the galaxy density map or CMB lensing potential is given by $W^x_L=C^{\grad x}_L/\hC^{xx}_L$ where $\hC^{xx}$ is the observed power spectrum of the measured mass tracer fields. In the galaxy delensing, the observed galaxy power spectrum contains the shot noise in the measurement of the galaxy number density. In the CMB lensing potential, the observed power spectrum is computed as the sum of the lensing potential spectrum and Gaussian reconstruction noise. 

In computing the lensing template from the galaxy density map, we use the galaxy density fields and $E$ modes up to $\l=2048$. For the CMB internal delensing, we reconstruct the lensing potential using $\Theta\Theta$ or $EB$ quadratic estimator \cite{Hu:2001kj,Okamoto:2002ik}. The CMB multipoles between $\l=500$ and $2048$ are used for the reconstruction without the CMB instrumental noise. 

%////////////////////////////////////////////////////////////////////////////////////////////////////%
\section{Simulation} \label{sim}
%////////////////////////////////////////////////////////////////////////////////////////////////////%

%<><><><><><><><><><><><><><><><><><><><><><><><><><><><><><><><><><><><><><><><><><><><><><><><><><>%
\begin{figure}
\bc
\includegraphics[width=89mm,clip]{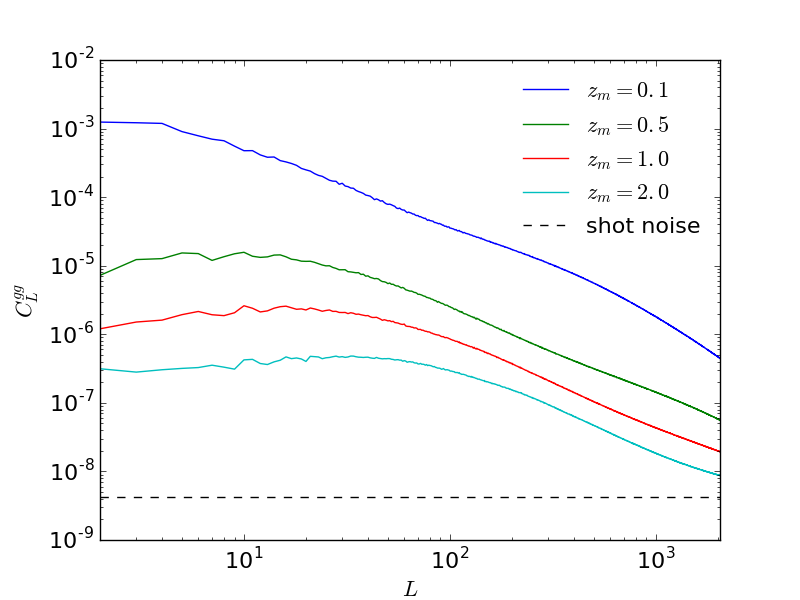}
\caption{
The galaxy power spectrum with varying the mean redshift of galaxies, $z_{\rm m}=0.1$, $0.5$ $1.0$ and $2.0$. The power spectrum is obtained from the simulations with the nonlinear growth. 
}
\label{fig:clgg}
\ec
\end{figure}
%<><><><><><><><><><><><><><><><><><><><><><><><><><><><><><><><><><><><><><><><><><><><><><><><><><>%

%<><><><><><><><><><><><><><><><><><><><><><><><><><><><><><><><><><><><><><><><><><><><><><><><><><>%
\begin{figure*}
\bc
\includegraphics[width=21cm,clip]{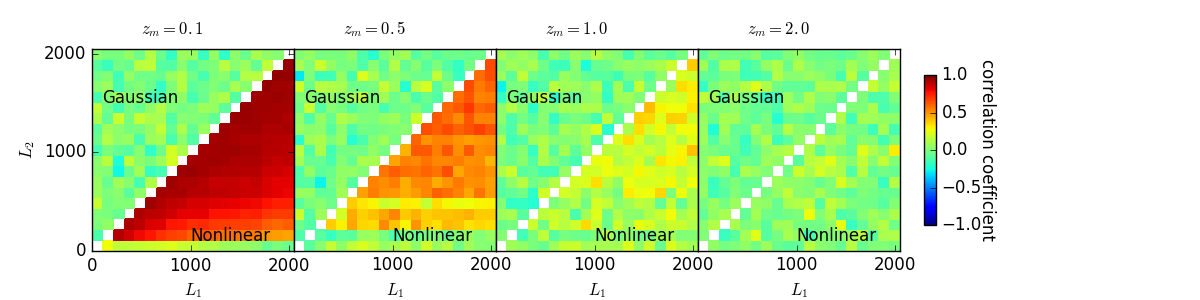}
\caption{
The correlation coefficients of the galaxy power spectrum with varying the mean redshift of galaxies, $z_{\rm m}=0.1$, $0.5$ $1.0$ and $2.0$. We show the cases with the random Gaussian simulation (upper left) or the simulation with the nonlinear growth (lower right). 
}
\label{fig:covgg}
\ec
\end{figure*}
%<><><><><><><><><><><><><><><><><><><><><><><><><><><><><><><><><><><><><><><><><><><><><><><><><><>%

%<><><><><><><><><><><><><><><><><><><><><><><><><><><><><><><><><><><><><><><><><><><><><><><><><><>%
\begin{figure}
\bc
\includegraphics[width=89mm,clip]{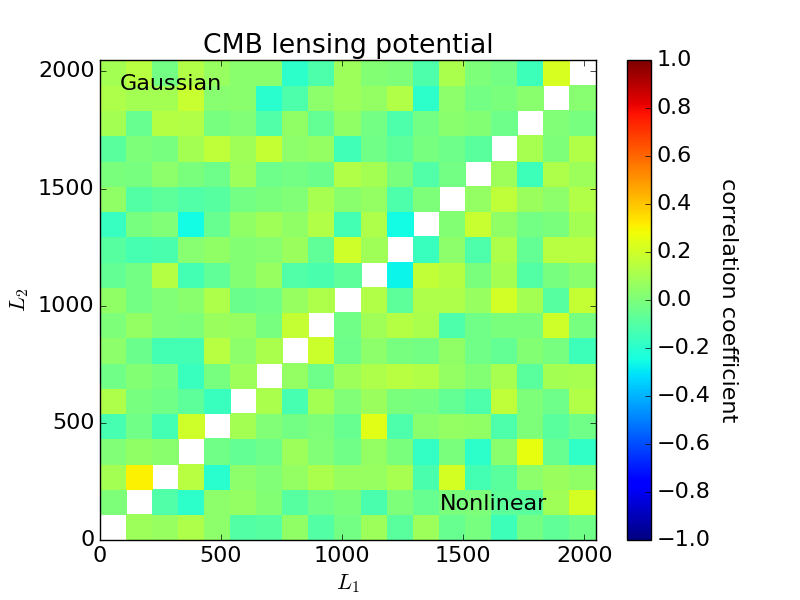}
\caption{
The correlation coefficients of the CMB lensing potential power spectrum.
}
\label{fig:covpp}
\ec
\end{figure}
%<><><><><><><><><><><><><><><><><><><><><><><><><><><><><><><><><><><><><><><><><><><><><><><><><><>%

Here, we describe our method of the $N$-body and ray-tracing simulation\tred{s} and how to simulate the lensed CMB maps and galaxy distribution. 

We first provide a short summary of the $N$-body and ray-tracing simulations. %used for constructing the foreground galaxies and lensed CMB maps, respectively. 
For detailed description, we refer readers to our previous paper~\cite{Takahashi:2017}. We performed cosmological $N$-body simulations of dark matter to reproduce the inhomogeneous mass distribution in the universe, which naturally includes the non-Gaussian fluctuations. Here, we followed the gravitational evolution of $2048^3$ particles in cubic simulation boxes by using a public code Gadget2~\cite{Springel:2001,Springel:2005}. The cubic boxes have side lengths of $450, \, 900, \, 1350,\dots, 6300 \, h^{-1} {\rm Mpc}$ from low to high redshift $(z \leq 7.1)$. We chose an arbitrary point in the box as the observer's position and constructed spherical lens shells with a thickness of $150h^{-1} \, {\rm Mpc}$ around the observer (which locates at the center of shells). \tred{Median radius of the $i$-th shell is $150 \times (i+1/2) \, h^{-1}{\rm Mpc}$ ($i=0,1,2,\dots$) in comoving scale. The inner shell is taken from the smaller box at lower redshift.} Then, we projected $N$-body particles onto each lens shell, and calculated surface mass density and deflection angle. For higher redshifts ($z=7.1-1100$), we prepared the lens shells based on the Gaussian fluctuations (instead of $N$-body method), which is a good approximation in the linear regime. Light rays, emitted from the observer, are deflected at each lens shell, and these ray paths are numerically evaluated up to the last scattering surface. Therefore, the simulation includes the post-Born correction. We used a public code {\tt GRayTrix}\footnote{\url{http://th.nao.ac.jp/MEMBER/hamanatk/GRayTrix/}} \cite{Hamana:2015,Shirasaki:2015} for this ray-tracing computation. This code also calculate the convergence, shear and rotation. The code relies on the {\tt Healpix} scheme \cite{Gorski:2004by}, and we adopt the angular resolution of $N_{\rm side}=4096$ throughout this paper. Then, we numerically obtained the deflection angle $\bm{d}(\hatn)$ on the full sky maps. We prepared $104$ such maps in total. % to be consistent with the main body text
We checked that the power spectra of lensing potential agree with the theoretical prediction calculated by {\tt CAMB} \cite{Lewis:1999bs} within $5\%$ up to $\l=3000$ (see Section 3.5 of Ref.~\cite{Takahashi:2017}). 

% CMB map
The lensed CMB maps are generated as follows. We first create unlensed Gaussian CMB maps which do not contain primary $B$ modes, using the CMB temperature and $E$-mode angular power spectra computed by {\tt CAMB}. Then, we use the above CMB deflection angles to remap the unlensed CMB temperature and polarization maps. We do not include the CMB instrumental noise in the lensed CMB maps. 

% galaxy map
Using the above $N$-body simulation, we also generate $104$ full-sky maps of the galaxies distribution. We populate ``galaxies'' as biased tracers of density contrast on the lens shells, based on given bias parameter $b(z)$ and radial number density distribution $dn/dz(z)$. We assume a simple scale-independent bias, $b(z)=\sqrt{1+z}$ \cite{Rassat:2008}. The number density of galaxies is $20$ [arcmin$^\tred{-2}$]. The normalized galaxy redshift distribution is assumed to be (e.g. \cite{Namikawa:2010})
%----------------------------------------------------------------------------------------------------%
\al{
	N(z) = \frac{\beta}{z_0\Gamma[(1+\alpha)/\beta]}
		\left(\frac{z}{z_0}\right)^\alpha \exp\left[-\left(\frac{z}{z_0}\right)^\beta\right]
	\,,
}
%----------------------------------------------------------------------------------------------------%
where the parameter, $z_0$, is determined by the mean redshift of the galaxies, $z_{\rm m}$: 
%----------------------------------------------------------------------------------------------------%
\al{
	z_0 = z_{\rm m}\frac{\Gamma[(1+\alpha)/\beta]}{\Gamma[(2+\alpha)/\beta]} 
	\,.
}
%----------------------------------------------------------------------------------------------------%
We choose $\alpha=2$ and $\beta=1$. We vary the mean redshift, $z_m$, from $0.1$ to $2.0$. 
\tred{The galaxy distribution with $z_m=1$ mimics a galaxy sample in the ongoig Subaru Hyper Suprime-Cam survey \cite{Aihara:2018}.}

% Gaussian map
In addition to the above maps, we also generate random Gaussian fields of galaxy distribution \tred{and lensing potential} whose power spectr\tred{a} coincide with that obtained from the simulation\tred{s} with the nonlinear growth. 
%We also generate random Gaussian fields of the lensing potential using the power spectrum obtained from the ray-tracing simulation. 
The lensing potential is generated so that the correlation between the lensing potential and galaxy distribution coincides with that of the simulation with the nonlinear growth. Then, the lensed CMB maps produced by using the Gaussian lensing potential are generated. 

% Check cl and covariance
Fig.~\ref{fig:clgg} shows the angular power spectrum of the galaxy distribution at each redshift measured from the simulation with the nonlinear growth, compared with the shot noise spectrum. The measured power spectra are dominated by the signal. Fig.~\ref{fig:covgg} shows the correlation coefficients of the angular power spectrum, i.e., 
%----------------------------------------------------------------------------------------------------%
\al{
	R_{bb'} = \frac{\cov_{bb'}}{\sqrt{\cov_{bb}\cov_{b'b'}}} \,,
}
%----------------------------------------------------------------------------------------------------%
where $\cov_b$ is the covariance of the binned angular spectrum defined as
%----------------------------------------------------------------------------------------------------%
\al{
	\cov_{bb'} = \ave{C^{gg}_b C^{gg}_{b'}} - \ave{C^{gg}_b}\ave{C^{gg}_{b'}} \,,
}
%----------------------------------------------------------------------------------------------------%
and $\ave{\cdots}$ denotes the average over $104$ realizations of our simulation. The correlation coefficients become significant at lower redshifts because the nonlinear growth of LSS leads to a non-Gaussianity in the density fluctuations. At large scales ($L\alt 100$), the nonlinear growth in the galaxy distribution decreases and the off-diagonal elements become zero. Even at $z_{\rm m}=1$, the correlation becomes $\sim 0.3$ at $L_1\sim L_2\agt 1000$. The galaxy distribution at $z_{\rm m}\alt 1$ is highly non-Gaussian. 

On the other hand, Fig.~\ref{fig:covpp} shows the correlation coefficients of the CMB lensing potential power spectrum. The correlation coefficients of the lensing power spectrum are not significant as the gravitational potential at high redshift, $z\sim 2$, is the dominant source of the CMB lensing effect.

%////////////////////////////////////////////////////////////////////////////////////////////////////%
\section{Results} \label{result}
%////////////////////////////////////////////////////////////////////////////////////////////////////%

The difference of the delensing results between the Gaussian and nonlinear-growth simulations appear in the angular power spectrum and its covariance. The discrepancy in the delensed $B$-mode spectrum leads to a bias in the estimate of the primary $B$ mode contributions. The increase of the covariance degrades the delensing efficiency and signal-to-noise \tred{ratio} of the primary $B$ modes. In the following sections, we summarize the discrepancy of the results between the Gaussian and nonlinear simulations. 

%++++++++++++++++++++++++++++++++++++++++++++++++++++++++++++++++++++++++++++++++++++++++++++++++++++%
\subsection{Delensed B-mode power spectrum}
%++++++++++++++++++++++++++++++++++++++++++++++++++++++++++++++++++++++++++++++++++++++++++++++++++++%

%<><><><><><><><><><><><><><><><><><><><><><><><><><><><><><><><><><><><><><><><><><><><><><><><><><>%
\begin{figure}
\bc
\includegraphics[width=89mm,clip]{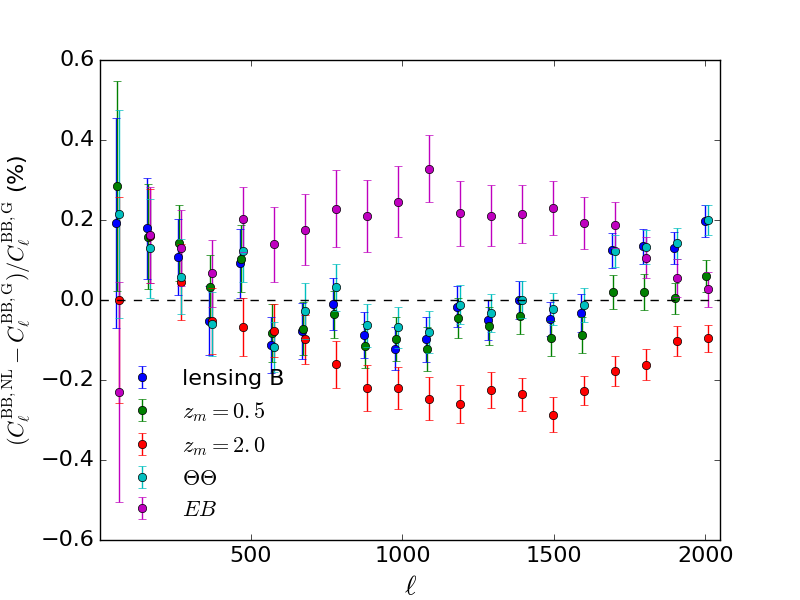}
\caption{
Fractional difference of the lensing or delensed $B$-mode spectra between the Gaussian and nonlinear-growth simulations. The delensing of $B$ mode is performed using a simulated galaxy distribution with the mean redshift of $z_{\rm m}=0.5$ or $2.0$, or a reconstructed CMB lensing map from $\Theta\Theta$ / $EB$ quadratic estimator. 
}
\label{fig:bb-diff}
\ec
\end{figure}
%<><><><><><><><><><><><><><><><><><><><><><><><><><><><><><><><><><><><><><><><><><><><><><><><><><>%

Fig.~\ref{fig:bb-diff} shows the fractional difference of the lensing or delensed $B$-mode power spectra between the Gaussian and nonlinear-growth simulations. As shown in Fig.~1 of Ref.~\cite{Lewis:Pratten:2016}, the nonlinear growth of the LSS produces additional sub-percent contributions in the lensing $B$-mode power spectrum. Our result is consistent with Ref.~\cite{Lewis:Pratten:2016}. 

The nonlinear growth also leads to additional contributions in the delensed $B$-mode spectra. The fractional change to the delensed $B$-mode spectrum with a mass tracer of $z_{\rm m}=0.5$ or $\T\T$ quadratic estimator is similar to that to the lensing $B$ mode. This is because the delensing efficiency is not so high (see Fig.~\ref{fig:resbb} below), and the nonlinear contributions in the delensed $B$ mode is not significantly different from that in the lensing $B$ mode. On the other hand, if we use the galaxies with $z_{\rm m}=2.0$, the fractional difference becomes sub-percent and is more significant than that in the case with $z_{\rm m}=0.5$ or the $\T\T$ estimator. This indicates that the nonlinear structure is not effectively removed even if the delensing efficiency is improved. 
Finally, if we use the reconstructed lensing map from the $EB$ estimator, the fractional difference becomes positive. Since the delensing efficiency using the $EB$ estimator is much better than that in the other cases, other nonlinear terms would remain in the fractional difference. In addition, the bias comes from the correlation between the $B$ modes to be delensed and in the $EB$ estimator \cite{Teng:2011xc,Namikawa:2017a}). This correlation is approximately proportional to the lensing $B$ mode, and leads to the additional nonlinear contribution in the fractional difference. 
%If we use the reconstructed lensing map from the $EB$ estimator, the fractional difference becomes positive. This further bias partially comes from the delensing method we used, i.e., making a lensing $B$-mode template which includes $\grad$ up to the linear order. The lack of $\mC{O}(\grad^2)$ terms in the lensing $B$-mode template leaves additional nonlinear terms, e.g., $\grad^3$ contributions arising from the correlation between the template and lensing $B$-modes, which are not involved in the original lensing $B$-mode. Since the delensing efficiency using the $EB$ estimator is much better than that in the other cases, the above effects become significant. Another significant source of the bias in the $EB$ case is the correlation between the $B$ modes to be delensed and in the $EB$ estimator \cite{Teng:2011xc,Namikawa:2017a}). This correlation is approximately proportional to the lensing $B$ mode, and leads to the additional nonlinear contribution in the fractional difference. 
%While a thorough explanation of the fractional difference requires analytic calculations of the delensed $BB$ spectrum including nonlinear contributions of the lensing potential, 
As described above, while the behavior of the nonlinear contribution in the $B$-mode spectra depends on the mass tracer, the nonlinear contribution in the lensing/delensed $BB$ spectrum is not significant compared to the statistical error expected in the ongoing and near future CMB experiments.

%++++++++++++++++++++++++++++++++++++++++++++++++++++++++++++++++++++++++++++++++++++++++++++++++++++%
\subsection{Correlation coefficients of power spectrum}
%++++++++++++++++++++++++++++++++++++++++++++++++++++++++++++++++++++++++++++++++++++++++++++++++++++%

%<><><><><><><><><><><><><><><><><><><><><><><><><><><><><><><><><><><><><><><><><><><><><><><><><><>%
%\begin{figure}
%\bc
%\includegraphics[width=89mm,clip]{fig_vbb_ratio.png}
%\caption{
%The variance of the delensed or lensing $B$-mode power spectrum (divided by the Gaussian case). 
%}
%\label{fig:vbb}
%\ec
%\end{figure}
%<><><><><><><><><><><><><><><><><><><><><><><><><><><><><><><><><><><><><><><><><><><><><><><><><><>%

%Next we discuss the variance of the delensed $B$-mode spectrum. Fig.~\ref{fig:vbb} plots the ratio of the variance of the delensed or lensing $B$-mode spectrum \tred{in} the nonlinear-growth simulation to that in the Gaussian simulation. Because the Monte Carlo error in our simulation is roughly $10\%$, the variance in the nonlinear-growth simulation is not significantly different from that in the Gaussian simulation. 

%<><><><><><><><><><><><><><><><><><><><><><><><><><><><><><><><><><><><><><><><><><><><><><><><><><>%
\begin{figure}
\bc
\includegraphics[width=89mm,clip]{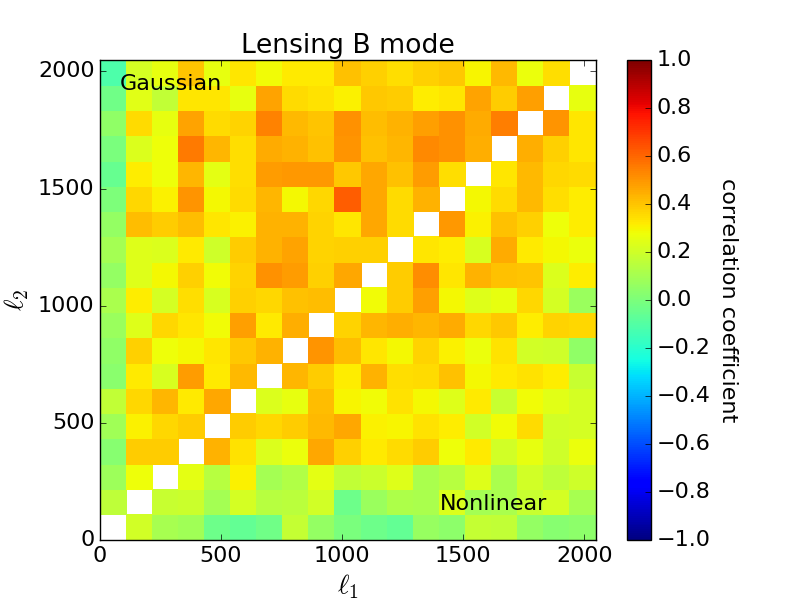}
\caption{
The correlation coefficients of the lensing $B$-mode power spectrum. 
}
\label{fig:covbb-lens}
\ec
\end{figure}
%<><><><><><><><><><><><><><><><><><><><><><><><><><><><><><><><><><><><><><><><><><><><><><><><><><>%

%<><><><><><><><><><><><><><><><><><><><><><><><><><><><><><><><><><><><><><><><><><><><><><><><><><>%
\begin{figure*}
\bc
\includegraphics[width=21cm,clip]{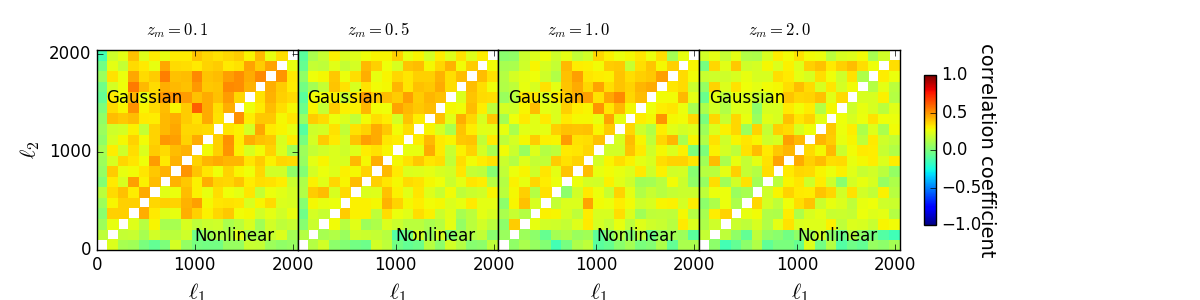}
\includegraphics[width=21cm,clip]{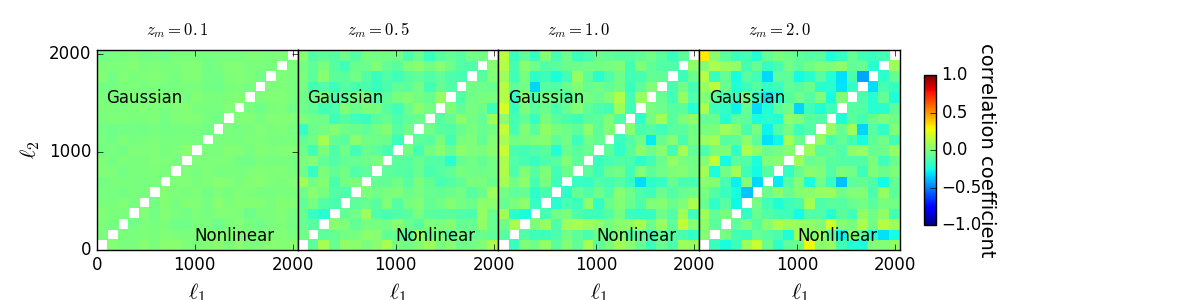}
\caption{
{\bf Top}: The correlation coefficients of the delensed $B$-mode power spectrum using galaxy distribution at $z_m=0.1$, $z_m=0.5$, $z_m=1.0$ and $z_m=2.0$ from left to right. 
{\bf Bottom}: The difference of the correlation coefficients between the delensed and lensing $B$-mode spectrum. 
}
\label{fig:covbb}
\ec
\end{figure*}
%<><><><><><><><><><><><><><><><><><><><><><><><><><><><><><><><><><><><><><><><><><><><><><><><><><>%

Fig.~\ref{fig:covbb-lens} shows the correlation coefficients of the lensing $B$-mode spectrum (the $B$-mode spectrum without delensing). As shown in \cite{BenoitLevy:2012va}, the non-Gaussianity becomes significant at smaller scales. Our results show that the nonlinear growth effect does not significantly change the correlations. 

Fig.~\ref{fig:covbb} shows the correlation coefficients of the delensed $B$-mode power spectrum, and the difference of the coefficients between the delensed and lensing $B$-mode spectra. In all cases, the nonlinear growth does not significantly affect the correlation coefficients of the delensed $B$-mode spectrum. 

\mf{
We also check that the change of the variance of the lensing/delensed $B$-mode spectrum by the inclusion of the nonlinear growth is also not significant. 
}

%<><><><><><><><><><><><><><><><><><><><><><><><><><><><><><><><><><><><><><><><><><><><><><><><><><>%
\begin{figure*}
\bc
\includegraphics[width=21cm,clip]{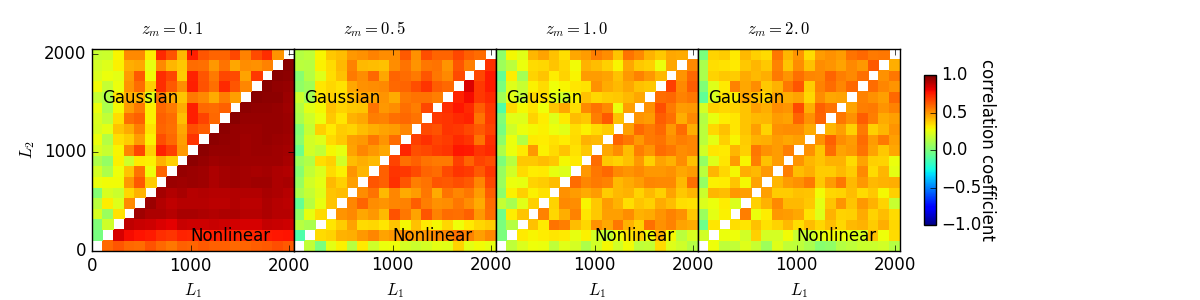}
\caption{
The correlation coefficients of the lensing template spectrum using galaxy distribution at $z_m=0.1$, $0.5$, $1.0$ and $2.0$. 
}
\label{fig:covbblt}
\ec
\end{figure*}
%<><><><><><><><><><><><><><><><><><><><><><><><><><><><><><><><><><><><><><><><><><><><><><><><><><>%

%<><><><><><><><><><><><><><><><><><><><><><><><><><><><><><><><><><><><><><><><><><><><><><><><><><>%
\begin{figure}
\bc
\includegraphics[width=89mm,clip]{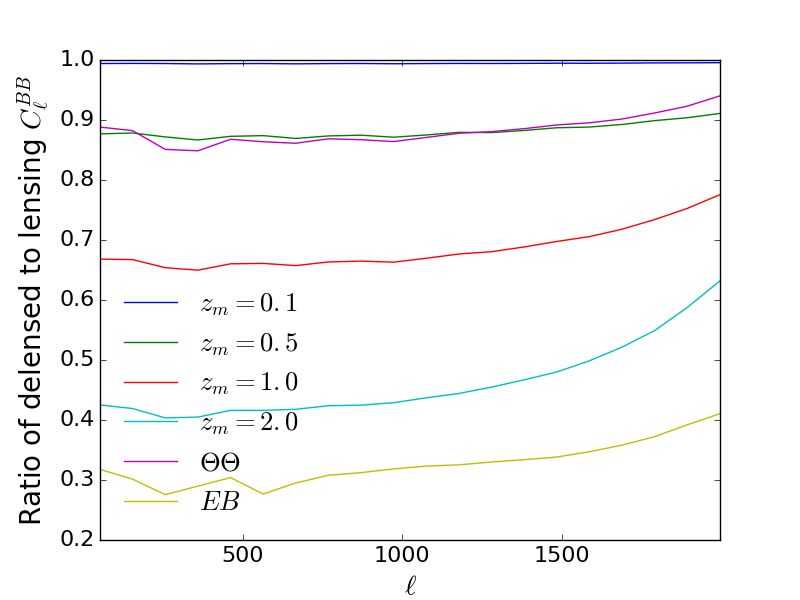}
\caption{
Ratio of the delensed to lensing $B$-mode power spectrum with varying the mean redshift of the galaxy survey. We also show the ratio with $\T\T$/$EB$ quadratic estimator. The simulation with the nonlinear growth is used. 
}
\label{fig:resbb}
\ec
\end{figure}
%<><><><><><><><><><><><><><><><><><><><><><><><><><><><><><><><><><><><><><><><><><><><><><><><><><>%

To see how the non-Gaussianity of the galaxy distribution shown in Fig.~\ref{fig:covgg} is suppressed, Fig.~\ref{fig:covbblt} shows the off-diagonal covariance of the lensing template given by \eq{Eq:delens-B}. As indicated by Fig.~\ref{fig:covgg}, the lensing template with mass tracers at low $z$ is also a non-Gaussian field and has non-negligible correlation coefficients of the power spectrum. \mf{This large correlation coefficients seen for low-$z$ templates make them sub-optimal estimators for the lensing B-modes shown in Fig.~\ref{fig:covbb}.} Even using low-$z$ mass tracers, however, the delensed $B$-mode spectrum is not significantly affected by the nonlinear growth. This is partly because the off-diagonal correlation matrix of the lensing template contributes to that of the delensed $B$ mode after scaled by, $\sqrt{f_bf_{b'}}$, with $f_b\equiv C_b^{\rm BB,temp}/C_b^{\rm BB,res}$ 
\footnote{The correlation coefficients of the delensed $B$-mode spectrum is given by $R^{\rm res}_{bb'}\simeq(\ave{C_b^{\rm BB,res}C_{b'}^{\rm BB,res}}-C_b^{\rm BB,res}C_{b'}^{\rm BB,res})/2\sqrt{C_b^{\rm BB,res}C_{b'}^{\rm BB,res}}$. This contains the correlation coefficients of the lensing template spectrum as $R^{\rm res}_{bb'}\supset (\ave{C_b^{\rm BB,temp}C_{b'}^{\rm BB,temp}}-C_b^{\rm BB,temp}C_{b'}^{\rm BB,temp})/2\sqrt{C_b^{\rm BB,\tred{res}}C_{b'}^{\rm BB,\tred{res}}}\simeq R_{bb'}^{\rm \tred{temp}}\sqrt{f_bf_{b'}}$ with $f_b=C_b^{\rm BB,temp}/C_b^{\rm BB,res}$.}. 
As shown in Fig.~\ref{fig:resbb}, the delensing efficiency decreases by decreasing the redshift of the mass tracers. For low-$z$ tracers, the scaling factor therefore becomes $f_b = (C_b^{\rm BB}-C_b^{\rm BB,res})/C_b^{\rm BB,res} \ll 1$. 

%<><><><><><><><><><><><><><><><><><><><><><><><><><><><><><><><><><><><><><><><><><><><><><><><><><>%
%\begin{figure}
%\bc
%\includegraphics[width=89mm,clip]{fig_vbb_ratio_zcmb.png}
%\caption{
%The variance of the delensed $B$-mode power spectrum (divided by the Gaussian case) using the CMB derived lensing potential. 
%}
%\label{fig:vbb_zcmb}
%\ec
%\end{figure}
%<><><><><><><><><><><><><><><><><><><><><><><><><><><><><><><><><><><><><><><><><><><><><><><><><><>%

%<><><><><><><><><><><><><><><><><><><><><><><><><><><><><><><><><><><><><><><><><><><><><><><><><><>%
\begin{figure*}
\bc
\includegraphics[width=89mm,clip]{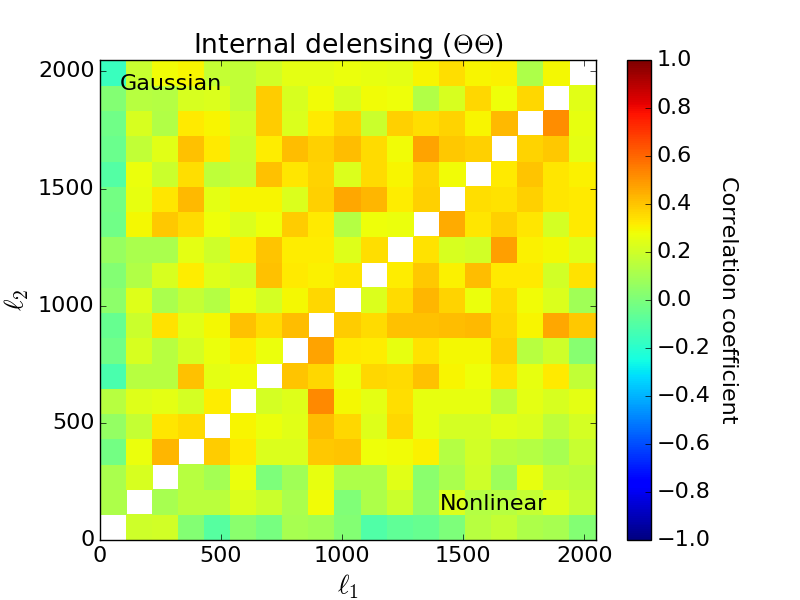}
\includegraphics[width=89mm,clip]{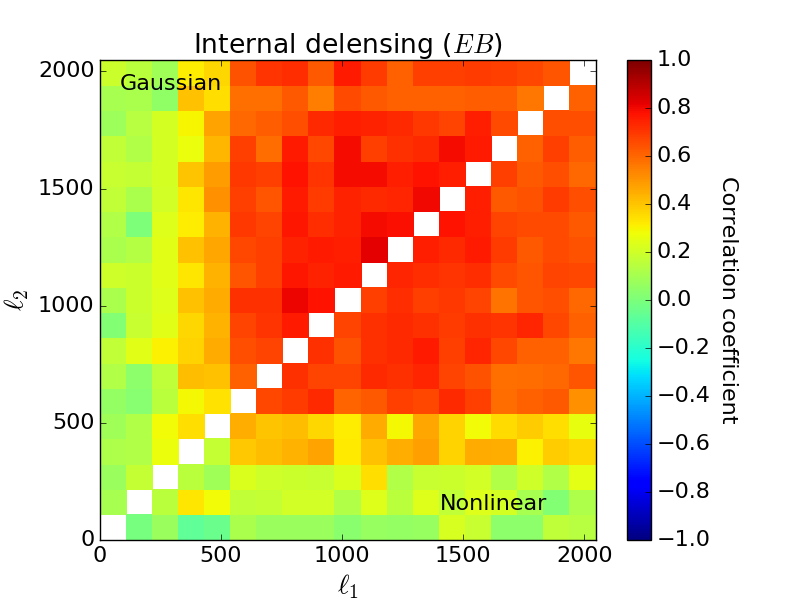}
\caption{
The correlation coefficients of the delensed $B$-mode spectrum with CMB derived lensing potential using $\Theta\Theta$ (Left) or $EB$ (Right) quadratic estimator. 
}
\label{fig:covbb_zcmb}
\ec
\end{figure*}
%<><><><><><><><><><><><><><><><><><><><><><><><><><><><><><><><><><><><><><><><><><><><><><><><><><>%

%Figs.~\ref{fig:vbb_zcmb} and ~\ref{fig:covbb_zcmb} show the variance and correlation coefficients of the delensed $B$-mode spectrum using the CMB derived lensing potential reconstructed using the $\Theta\Theta$ or $EB$ quadratic estimators. 
\mf{
Fig.~\ref{fig:covbb_zcmb} shows the correlation coefficients of the delensed $B$-mode spectrum using the CMB derived lensing potential reconstructed using the $\Theta\Theta$ or $EB$ quadratic estimators. Similar to the galaxy delensing case, the impact of the nonlinear growth is not significant. We check that the change of the variance of the delensed $B$-mode spectrum by the inclusion of the nonlinear growth is also not significant. Note that, while the impact of the nonlinear growth is not important, the simulation results indicate that the off-diagonal element of the covariance using $EB$ estimator has non-trivial non-Gaussian feature at smaller scales. One possibility of the origin of this feature is that the $B$ mode to be delensed correlates with the $B$ mode in the lensing reconstruction (see e.g. \cite{Namikawa:2017a}). The detailed study on the covariance of the delensed $B$ mode using $EB$ estimator is highly involved, and will be investigated in our future work. 
}

%++++++++++++++++++++++++++++++++++++++++++++++++++++++++++++++++++++++++++++++++++++++++++++++++++++%
\subsection{Power spectrum amplitude}
%++++++++++++++++++++++++++++++++++++++++++++++++++++++++++++++++++++++++++++++++++++++++++++++++++++%

%<><><><><><><><><><><><><><><><><><><><><><><><><><><><><><><><><><><><><><><><><><><><><><><><><><>%
% table
\begin{table}
\bc
\caption{
The variance of the power spectrum amplitude, $\sigma(A)\times 10^3$, using $104$ realizations of CMB/galaxy maps, with $20$ multipole bins between $2\leq L\leq 2048$. 
}
\label{Table:amp} \vs{0.5}
\begin{tabular}{lcc} \hline
 & \tred{G}auss & \tred{N}onlinear \\ \hline 
Lensing $BB$ & 1.89 & 1.91 \\ 
$z_m=0.1$ & 1.89 & 1.92 \\ 
$z_m=0.5$ & 1.82 & 1.78 \\ 
$z_m=1.0$ & 1.86 & 1.60 \\ 
$z_m=2.0$ & 1.40 & 1.55 \\ \hline
$\Theta\Theta$ & 1.48 & 1.55 \\
$EB$ & 1.86 & 1.92 \\ \hline
\end{tabular}
\ec
\end{table}
%<><><><><><><><><><><><><><><><><><><><><><><><><><><><><><><><><><><><><><><><><><><><><><><><><><>%

To quantify the impact of the nonlinear growth on the delensed $B$-mode spectrum, we evaluate the statistical error of the amplitude of the angular power spectrum (i.e. the inverse of the signal-to-noise of the power spectrum). The amplitude is estimated from (e.g., \cite{Bicep2KeckArray:2016,Bicep2KeckArray:2017})
%----------------------------------------------------------------------------------------------------%
\al{
	A = \frac{1}{\sum_b w_b} \sum_b w_b \frac{\widehat{C}_b}{C_b^{\rm ref}} 
    \,,
}
%----------------------------------------------------------------------------------------------------%
with $w_b = \sum_{b'} C_b^{\rm ref}{\rm Cov}^{-1}_{bb'}C_{b'}^{\rm ref}$. If the off-diagonal elements of the correlation coefficients are significant, the statistical error of the above amplitude increases. 

Table \ref{Table:amp} shows the variance of the power spectrum amplitude from $104$ realizations. The covariance and $C_b^{\rm ref}$ are computed from the simulation. Even in the Gaussian cases, the off-diagonal elements of the correlation coefficients exist due to the non-Gaussianity of the lensing $B$ mode. The variance decreases as the source redshift increases. This indicates that the delensing removes lensing contributions instead of adding additional nonlinear growth effects, and the off-diagonal covariance is suppressed. Within the simulation error, there are no significant difference between the results from the Gaussian and nonlinear growth simulations.

%////////////////////////////////////////////////////////////////////////////////////////////////////%
\section{Summary and discussion} \label{summary}
%////////////////////////////////////////////////////////////////////////////////////////////////////%

% summary
We explored the impact of the nonlinear growth of the LSS on delensing. We considered delensing using the galaxy distribution or the CMB-derived lensing potential. We found that the corrections to the delensed $B$-mode spectrum are $\sim 0.3$\% at $\l=1000-2000$ with high-$z$ mass tracers ($z_{\rm m}\sim 2$). While the effect of the nonlinear growth is significant on the lensing template with low-$z$ mass tracers, the correlation coefficients of the delensed $B$-mode spectrum are not significantly affected by the nonlinear growth. The impact of the nonlinear growth is also not significant on the correlation coefficients of the delensed $B$-mode spectrum with the CMB-derived lensing potential. 

% discussion
We ignored the CMB instrumental noise in the simulation. If the CMB noise is nearly Gaussian, the contribution from the noise does not create non-Gaussianity in the delensed $B$ mode and our conclusion above is unchanged. 

We focused on the galaxy distribution as a mass-tracer for delensing. The CIB is another interesting observable for the CMB delensing \cite{Simard:2015,Sherwin:2015}. The conclusion would be, however, also the same even if we use other mass tracers such as CIB because the signal part of mass tracers comes from the same underlying density fields. For example, in the CIB case, the source mostly comes from the density fluctuations at higher redshifts ($z\agt 1$) (e.g. \cite{P13:CIB}), and the results would be similar to the case with $z_{\rm m}\sim 1-2$ in our simulation. 

% public data
The lensed CMB maps, foreground matter distribution and a Fortran code to make mock galaxy distribution are publicly available. If you want to download them, please visit our website (\url{http://cosmo.phys.hirosaki-u.ac.jp/takahasi/allsky_raytracing/}).

% BACK MATTER 
% Acknowledgments %
\begin{acknowledgments}
We thank Vanessa Boehm, Yuji Chinone, Akito Kusaka, Alessandro Manzotti and Hironao Miyatake for helpful comments. TN acknowledges the support from the Ministry of Science and Technology (MOST), Taiwan, R.O.C. through the MOST research project grants (no. 107-2112-M-002-002-MY3). RT acknowledges the support from Grant-in-Aid for Scientific Research from the JSPS Promotion of Science (No. 17H01131) and MEXT Grant-in-Aid for Scientific Research on Innovative Areas (No. 15H05893). This research used resources of the National Energy Research Scientific Computing Center, which is supported by the Office of Science of the U.S. Department of Energy under Contract No. DE-AC02-05CH11231.
Numerical computations were in part carried out on Cray XC30 at Center for Computational Astrophysics, National Astronomical Observatory of Japan.
\end{acknowledgments}

% Appendix %
%\onecolumngrid
\appendix

%////////////////////////////////////////////////////////////////////////////////////////////////////%
\section{Delensing analysis for cut-sky} \label{appA}
%////////////////////////////////////////////////////////////////////////////////////////////////////%

The main motivation of this paper is to see the impact of the nonlinear growth in the delensing analysis. However, in practical analysis, many other possible effects should be simultaneously taken into account. One of the most significant effects in the delensing analysis would be the survey boundary due to finite sky coverage and Galactic/point-source masking. In this section, we show some results of delensing using finite size of CMB and galaxy density maps. 

%++++++++++++++++++++++++++++++++++++++++++++++++++++++++++++++++++++++++++++++++++++++++++++++++++++%
\subsection{Analysis in cut-sky}
%++++++++++++++++++++++++++++++++++++++++++++++++++++++++++++++++++++++++++++++++++++++++++++++++++++%

Since the difference of the results between the Gaussian and nonlinear simulations is very small, we only use the nonlinear-growth simulation in the cut-sky analysis. We cut $5\times 5$ deg$^2$ region from each realization of the lensed CMB map and galaxy density map at $z_{\rm m}=1$. The size of this patch is chosen to mimic the overlap region between the Subaru-Hyper Suprime Cam \cite{HSC:white-paper} and Polarbear \cite{PB1,PB2}. These maps are then multiplied by a simple analytic apodization function and transformed to Fourier space. For the lensed CMB maps, we reduce $E$-to-$B$ leakage by applying the $\chi$-field estimator \cite{Smith:2005:chi-estimator}. The delensing in flat sky is performed by first constructing the lensing template from the cut-sky $E$ mode and galaxy density, and then removing it from the observed $B$ mode. We simply apply the full-sky diagonal Wiener filter for $E$ mode and galaxy density fields in constructing the lensing template in flat sky. This simplification could make the delensing process sub optimal, but this approach is computationally easy for the implementation. The multipole range of the $E$ mode and galaxy density fields is the same as that in the full-sky analysis. We ignore the CMB instrumental noise. 

%++++++++++++++++++++++++++++++++++++++++++++++++++++++++++++++++++++++++++++++++++++++++++++++++++++%
\subsection{Delensed $B$-mode spectrum and its correlation coefficient}
%++++++++++++++++++++++++++++++++++++++++++++++++++++++++++++++++++++++++++++++++++++++++++++++++++++%

%<><><><><><><><><><><><><><><><><><><><><><><><><><><><><><><><><><><><><><><><><><><><><><><><><><>%
\begin{figure}
\bc
\includegraphics[width=89mm,clip]{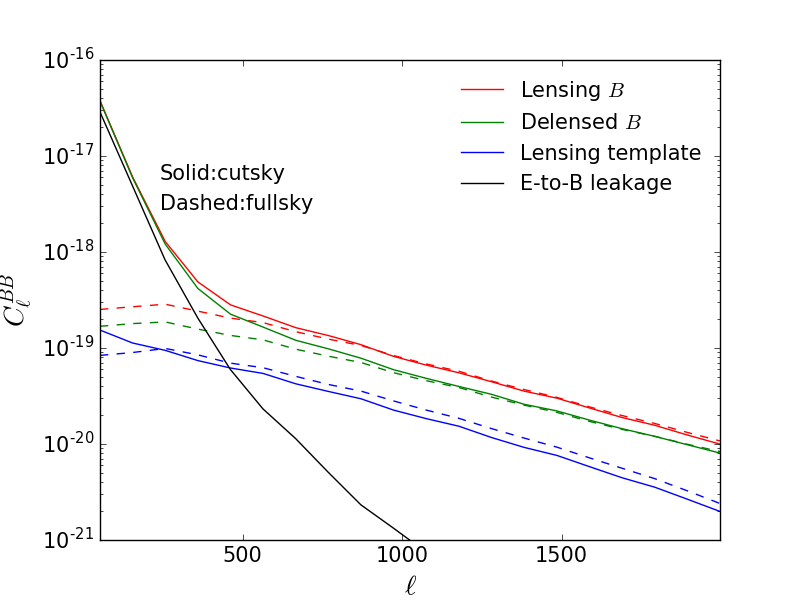}
\caption{
The lensing and delensed $B$-mode spectrum from the $5\times 5$ deg$^2$ cut-sky map. The delensing uses the galaxy number density with $z_{\rm m}=1$. 
}
\label{fig:cut:bb}
\ec
\end{figure}
%<><><><><><><><><><><><><><><><><><><><><><><><><><><><><><><><><><><><><><><><><><><><><><><><><><>%

%<><><><><><><><><><><><><><><><><><><><><><><><><><><><><><><><><><><><><><><><><><><><><><><><><><>%
\begin{figure}
\bc
\includegraphics[width=89mm,clip]{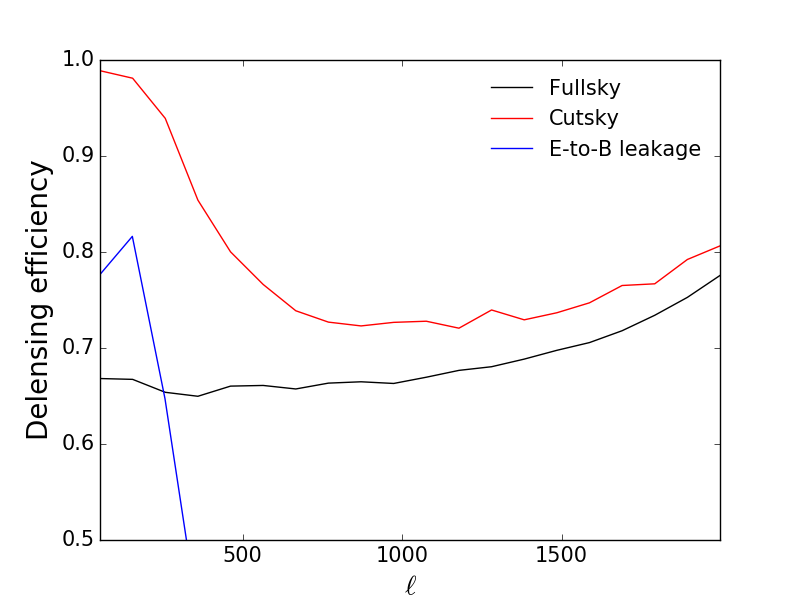}
\caption{
The delensing efficiency for the delensing using the cut-sky map. 
}
\label{fig:cut:bbeff}
\ec
\end{figure}
%<><><><><><><><><><><><><><><><><><><><><><><><><><><><><><><><><><><><><><><><><><><><><><><><><><>%

%<><><><><><><><><><><><><><><><><><><><><><><><><><><><><><><><><><><><><><><><><><><><><><><><><><>%
\begin{figure}
\bc
\includegraphics[width=89mm,clip]{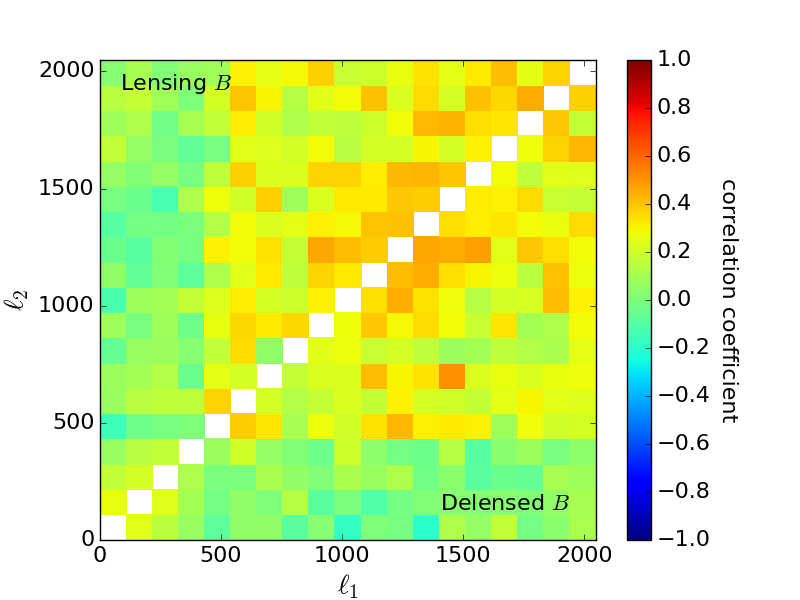}
\caption{
The correlation coefficients of the lensing and delensed $B$-mode spectrum from the cut-sky map.
}
\label{fig:cut:covbb}
\ec
\end{figure}
%<><><><><><><><><><><><><><><><><><><><><><><><><><><><><><><><><><><><><><><><><><><><><><><><><><>%

%<><><><><><><><><><><><><><><><><><><><><><><><><><><><><><><><><><><><><><><><><><><><><><><><><><>%
\begin{figure}
\bc
\includegraphics[width=89mm,clip]{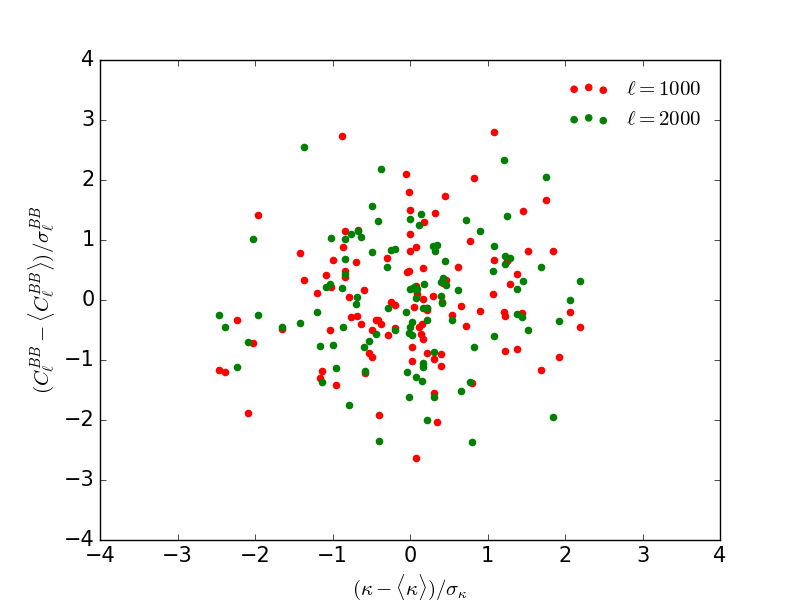}
\caption{
Correlation between the mean convergence and lensing $B$-mode spectrum at the $5\times 5$ deg$^2$ patch. The operation, $\ave{\cdots}$, is the average over the simulation realizations. 
}
\label{fig:cut:ssc}
\ec
\end{figure}
%<><><><><><><><><><><><><><><><><><><><><><><><><><><><><><><><><><><><><><><><><><><><><><><><><><>%

Fig.~\ref{fig:cut:bb} shows the lensing and delensed $B$-mode spectra in cut-sky analysis, compared with those obtained from the full-sky analysis. The $E$-to-$B$ leakage is obtained by removing the $B$ mode before the map is cut. At large scale ($\l<500$), the $B$-mode spectrum is dominated by the leakage from the $E$ mode due to the survey boundary and sky projection effect. The lensing template in cut-sky is slightly suppressed compared to that in full-sky. 

Fig.~\ref{fig:cut:bbeff} shows the delensing efficiency, i.e., the ratio of the delensed (or $E$-to-$B$ leakage) to lensing $B$-mode spectrum. In cut-sky analysis, the delensing efficiency reduces by $\sim 5$\% at small scales ($\l\geq 1000$). At large scales ($\l<500$), the $E$-to-$B$ leakage dominates the $B$ mode, and this contribution should be removed in addition to the lensing $B$ mode. One possible approach is to use the matrix method as discussed in the BICEP/Keck Array analysis \cite{BKVII}. Even if the matrix method is applied, the large-scale $B$ mode measurement from ground suffers from e.g. the 1/f noise, and the delensing of the $B$ mode at $\l<500$ is relatively not important for the Polarbear-like patch considered here. 

Fig.~\ref{fig:cut:covbb} shows the correlation coefficient of the $B$-mode spectrum in cut-sky. As shown in Fig.~\ref{fig:covbb}, in full sky, the delensing procedure reduces the correlation coefficients of the delensed $B$-mode spectrum. In contrast, the correlation coefficients are almost not decreased by the removal of the lensing $B$ mode in cut-sky. 

\mf{
Finally, in Fig.~\ref{fig:cut:ssc}, we show the correlation between the mean convergence in the small patch and the lensing $B$-mode power spectrum to see the effect of the super-survey mode \cite{TakadaHu:2013}. If the super-survey mode is significant, the power spectrum highly correlates with the mean convergence. We find that the correlation coefficient is at percent level and the super-survey mode is negligible. The analytic estimate of this effect is studied by Ref.~\cite{Manzotti:2014}, and the impact of the super-survey mode is shown to be negligible in the lensing $B$ mode spectrum. Our result is consistent with their results and justifies the approximation in their paper. }

% References %
\twocolumngrid
\bibliographystyle{mybst}
\bibliography{exp,main}

\end{document}